\documentclass[iop,twocolappendix,appendixfloats]{emulateapj}
\usepackage{apjfonts}
\usepackage{color}
\newcommand\beq{\begin{equation}}
\newcommand\eeq{\end{equation}}

\shorttitle{Inferring the halo mass of a galaxy}
\shortauthors{M. Oguri and Y.-T. Lin}

\begin{document}
%\begin{CJK*}{UTF8}{gbsn}

\title{Inferring Host Dark Matter Halo Masses 
of Individual Galaxies from Neighboring Galaxy Counts}

\author{
Masamune Oguri\altaffilmark{1,2,3} and
Yen-Ting Lin\altaffilmark{4,3}
}

\altaffiltext{1}{Research Center for the Early Universe, University of Tokyo, 7-3-1
  Hongo, Bunkyo-ku, Tokyo 113-0033, Japan; masamune.oguri@ipmu.jp}
\altaffiltext{2}{Department of Physics, University of Tokyo, 7-3-1 Hongo,
Bunkyo-ku, Tokyo 113-0033, Japan}
\altaffiltext{3}{Kavli Institute for the Physics and Mathematics of
  the Universe (Kavli IPMU, WPI), University of Tokyo, 5-1-5
  Kashiwanoha, Kashiwa, Chiba 277-8583, Japan} 
\altaffiltext{4}{Institute of Astronomy and Astrophysics, Academia Sinica, Taipei, Taiwan; ytl@asiaa.sinica.edu.tw}

%\end{CJK*}

%%%%%%%%%%%%%%%%%%%%%%%%%%%%%%%%%%%%%%%%%%%%%%
%%%%%%%%%%%%%%%%%%%%%%%%%%%%%%%%%%%%%%%%%%%%%%
\begin{abstract}
How well can we infer host dark matter halo masses of individual
galaxies? Based on the halo occupation distribution (HOD) framework,
we analytically compute the number of neighboring galaxies within a
cylinder of some redshift interval and radius in transverse comoving
distance. The result is used to derive the conditional probability
distribution function (PDF) of the host halo mass of a galaxy, given
the neighboring galaxy counts. We compare our analytic results with
those obtained using a realistic mock galaxy catalog, finding
reasonable agreements. We find the optimal cylinder radius to be 
$\sim 0.5-1h^{-1}{\rm Mpc}$ for the inference of halo masses. The PDF
is generally broad, and sometimes has two peaks at low- and high-mass
regimes, because of the effect of chance projection along the
line-of-sight. Potential applications and extensions of the new
theoretical framework developed herein are also discussed.
\end{abstract}


%%%%%%%%%%%%%%%%%%%%%%%%%%%%%%%%%%%%%%%%%%%%%%%%%%%%%%%%%%%%%%%%%%%%%%
\section{Introduction}
\label{sec:intro}
%%%%%%%%%%%%%%%%%%%%%%%%%%%%%%%%%%%%%%%%%%%%%%%%%%%%%%%%%%%%%%%%%%%%%%

Understanding the connection between galaxies and dark matter halos
is one of the most important goals of astrophysics and cosmology. The
framework of the halo occupation distribution (HOD) provides a
powerful means of studying the relation of galaxy distributions with
underlying dark matter distributions 
\citep{seljak00,peacock00,scoccimarro01,cooray02,kravtsov04,hamana04,zheng05}. 
The HOD characterizes the galaxy bias in terms of the halo occupation
number, the average number of galaxies within individual dark halos as
a function of halo masses, $\langle N(M)\rangle$, and is very
successful in explaining clustering properties of galaxies 
\citep[e.g.,][]{zehavi11,white11,wake11,leauthaud12,coupon12,hikage14,more14}. 
The idea of the HOD has also been applied to rarer cosmological
objects such as quasars \citep{kayo12,richardson12,shen13}.

The HOD model tells us how galaxies are connected to dark matter
halos {\it statistically}. A natural extension is then to consider how
well we can connect galaxies to dark matter halos for {\it individual}
galaxies. Naively we can infer the host halo property of individual
galaxies by examining number densities of galaxies around the target
galaxies, since a higher number density suggests that the target
galaxy is more likely to reside in a group- or cluster-scale dark
matter halo. However it is also possible that the galaxy actually
resides in a low-mass halo which appears to be associated with a
massive halo because of chance projection along the line-of-sight. The
redshift information is usually not sufficient to remove the chance
projection, particularly because the observed spatial distributions of
galaxies in massive halos are significantly elongated in the redshift
direction due to peculiar motions of the galaxies (the so-called
Fingers of God effect).  

In this paper, we develop a theoretical framework to compute the
probability distribution function (PDF) of host halo masses of
individual galaxies given number counts of nearby galaxies,
assuming that the true HOD is known a priori. We employ the HOD
formalism to predict, for each galaxy, the expected number of
neighboring galaxies within a cylinder of redshift interval $\pm
\Delta z$ and transverse comoving distance within $r_{\rm p,max}$. 
We include contributions  from both galaxies within the same host halo
and those residing in different halos. The result is used to derive 
the conditional PDF of host halo masses of a galaxy given the
neighboring galaxy counts within the cylinder
\citep[see also][for a similar study using a semi-analytic model
 of galaxy formation]{haas12}.
We compare our results
with those from a mock galaxy catalog constructed from the Millennium
Run simulation \citep{springel05,henriques12}.   

Neighboring galaxy counts, or sometimes referred as
counts-in-cylinders, have also been used to constrain the HOD from
observations. For instance, \citet{lin04} directly counted number of
galaxies within massive clusters to constrain the HOD at high mass 
end. \citet{chatterjee13} counted the number of quasars in galaxy
clusters to try to break degeneracies in the quasar HOD.
 \citet{reid09} conducted a counts-in-cylinders analysis to study
the HOD of luminous red galaxies \citep[see also][]{ho09}. 
In the previous studies, however, connecting galaxies to dark halos,
including the estimate of chance projection, has been attempted rather
empirically, e.g., using mock galaxy catalogs. The new theoretical
framework developed in this paper provides a more rigorous means of
connecting neighboring galaxy counts and the underlying HOD.

The structure of this paper is as follows. In Section~\ref{sec:nc}, we
present our analytic model to compute neighboring galaxy counts within
the HOD framework. Using the result, in Section~\ref{sec:pdf} we
derive conditional PDFs of the host halo mass given neighboring galaxy
counts. We compare our analytic model predictions with results from
the mock galaxy catalog in Section~\ref{sec:mock}. We discuss
possible extensions and applications of our model in
Section~\ref{sec:ext}, and give a summary in Section~\ref{sec:summary}.
Where necessary we use the $\Lambda$-dominated cold dark matter
cosmology model employed by the Millennium Run simulation, where
$\Omega_m=0.25$, $\Omega_\Lambda=0.75$, 
$H_0=100h~{\rm km~s^{-1}~Mpc^{-1}}$ with $h=0.73$, $n_s=1$, and
$\sigma_8=0.9$.  

%%%%%%%%%%%%%%%%%%%%%%%%%%%%%%%%%%%%%%%%%%%%%%%%%%%%%%%%%%%%%%%%%%%%%%
\section{Halo occupation distribution approach to neighboring galaxy counts}
\label{sec:nc}
%%%%%%%%%%%%%%%%%%%%%%%%%%%%%%%%%%%%%%%%%%%%%%%%%%%%%%%%%%%%%%%%%%%%%%

%%%%%%%%%%%%%%%%%%%%%%%%%%%%%%%%%%%%%%%%%%%%%%%%%%%%%%%%%%%%%%%%%%%%%%
\subsection{Halo occupation distribution}
%%%%%%%%%%%%%%%%%%%%%%%%%%%%%%%%%%%%%%%%%%%%%%%%%%%%%%%%%%%%%%%%%%%%%%

The HOD model specifies the mean halo occupation number
$\langle N(M)\rangle$ as a function of halo mass $M$. Given the mass
function and clustering properties of dark matter halos calibrated by
$N$-body simulations, the HOD modeling enables us to analytically
compute clustering properties of galaxies.  Following \citet{zheng05}, 
we consider central and satellite galaxies separately  
%%%%%%%%%%
\begin{equation}
\langle N(M)\rangle =\langle N_{\rm cen}(M)\rangle +
\langle N_{\rm sat}(M)\rangle,
\end{equation}
%%%%%%%%%%
where $\langle N_{\rm cen}(M)\rangle$ and $\langle N_{\rm
  sat}(M)\rangle$ describe the central and satellite components,
respectively. We parametrize both components as 
\citep[e.g.,][]{zheng05,white11}
%%%%%%%%%%
\begin{equation}
\langle N_{\rm cen}(M)\rangle=
\frac{1}{2}{\rm erfc}\left[\frac{\ln(M_{\rm min}/M)}
{\sqrt{2}\sigma}\right],
\label{eq:hod_cen}
\end{equation}
%%%%%%%%%%
%%%%%%%%%%
\begin{equation}
\langle N_{\rm sat}(M)\rangle=
\langle N_{\rm cen}(M)\rangle\left
(\frac{M-\kappa M_{\rm cut}}{M_1}\right)^\alpha.
\label{eq:hod_sat}
\end{equation}
%%%%%%%%%%
Here $M_{\rm min}$ is the characteristic minimum mass that can host
central galaxies, $\sigma$ is the characteristic transition width,
$\kappa M_{\rm cut}$ is the cutoff mass for satellites, and $M_1$ and
$\alpha$ determine how the mean number of satellite galaxies within
halos grows as a function of halo masses (also see Table~\ref{tab:hod}
below). This functional form is motivated by cosmological galaxy
formation simulations and is known to explain various observational
data very well. 

In the following, the galaxies for which we wish to infer host dark
matter halo masses are referred to as {\it target} galaxies, while the
galaxies surrounding them (as seen in projection on the sky) are called
{\it neighboring} galaxies. Quantities associated with target
(neighboring) galaxies are denoted with a subscript or superscript t
(n). The subscript p serves as a reminder when a quantity is evaluated
in projection. 

%%%%%%%%%%%%%%%%%%%%%%%%%%%%%%%%%%%%%%%%%%%%%%%%%%%%%%%%%%%%%%%%%%%%%%
\subsection{Neighboring galaxy counts}
%%%%%%%%%%%%%%%%%%%%%%%%%%%%%%%%%%%%%%%%%%%%%%%%%%%%%%%%%%%%%%%%%%%%%%

Our primary goal is to compute number counts of neighboring galaxies
around a target galaxy. Throughout the paper we count the number of
neighboring galaxies within a cylinder defined by a redshift interval
$\pm \Delta z$ and transverse comoving distance 
$r_{\rm p}<r_{\rm p, max}$. We denote these neighboring galaxy counts
as $N_{\rm p}=N_{\rm p}(<r_{\rm p, max})$. For a galaxy at redshift
$z$, the volume integral of the cylinder is explicitly written as 
%%%%%%%%%%
\begin{equation}
\int dV_{\rm c} =\int_{z-\Delta z}^{z+\Delta z} \frac{c\,dz}{H(z)}
\int_0^{r_{\rm p,max}} 2\pi r_{\rm p} dr_{\rm p}.
\end{equation}
%%%%%%%%%%

It is convenient to derive an explicit expression for the volume
integral of the two-point correlation function $\xi(\mathbf{x})$ over
the cylinder. Suppose the redshift interval is large enough to include
correlated structures, we have
%%%%%%%%%%
\begin{equation}
\int dV_{\rm c} \xi(\mathbf{x})=
\int_0^{r_{\rm p,max}} 2\pi r_{\rm p} dr_{\rm p} w_{\rm p}(r_{\rm p}).
\end{equation}
%%%%%%%%%%
The projected correlation function $w_{\rm p}(r_{\rm p})$ is related
to the power spectrum $P(k)$ via 
%%%%%%%%%%
\begin{equation}
w_{\rm p}(r_{\rm p})=\int \frac{k\,dk}{2\pi}P(k)J_0(kr_{\rm p}),
\end{equation}
%%%%%%%%%%
with $J_\alpha(x)$ being the Bessel functions of the first kind. 
Using the relation $\int_0^x dx'\, x'J_0(x')=xJ_1(x)$
we then obtain
%%%%%%%%%%
\begin{equation}
\int dV_{\rm c} \xi(\mathbf{x})= \int_0^\infty dk\,
r_{\rm p,max}P(k)J_1(kr_{\rm p,max}).
\label{eq:xi_pk}
\end{equation}
%%%%%%%%%%

%%%%%%%%%%%%%%%%%%%%%%%%%%%%%%%%%%%%%%%%%%%%%%%%%%%%%%%%%%%%%%%%%%%%%%
\subsubsection{1-halo term}\label{sec:1h}
%%%%%%%%%%%%%%%%%%%%%%%%%%%%%%%%%%%%%%%%%%%%%%%%%%%%%%%%%%%%%%%%%%%%%%

We first consider the so-called 1-halo term, i.e., number
counts of neighboring galaxies that reside in the same dark
halo. Neighboring galaxy counts should depend on where in the halo 
the target galaxy is located. We can include this effect by computing
neighboring galaxy counts for central and satellite galaxies
separately.  

Using the expression given in Equation~(\ref{eq:xi_pk}), we can derive
the average number of neighboring galaxies within the cylinder around a
central galaxy as 
%%%%%%%%%%
\begin{equation}
\langle N_{\rm p}^{\rm 1h,c}\rangle_M
=\int_0^\infty dk\, r_{\rm p,max}
J_1(kr_{\rm p,max})\langle N_{\rm sat}(M)\rangle \tilde{u}(k|M),
\end{equation}
%%%%%%%%%%
where $\tilde{u}(k|M)$ is the Fourier transform of the normalized
number density profile of satellite galaxies, $u(r|M)$
%%%%%%%%%%
\begin{equation}
\tilde{u}(k|M)=\int_0^{r_{\rm 200c}}4\pi r^2 dr\, u(r|M)\, j_0(kr),
\end{equation}
%%%%%%%%%%
with $j_0(x)=\sin(x)/x$ being the spherical Bessel function. In this
paper, we define the halo mass $M$ by 
the total mass within a sphere of $r_{\rm 200c}$, within which the mean
density is 200 times the critical density of the Universe. We use this
overdensity in order to compare our results with those from the mock
galaxy catalog of \citet[][see
  Section~\ref{sec:mock}]{henriques12}. We make the simplifying
assumption that the satellite number density profile follows the
matter density profile of dark matter 
\citep[][hereafter NFW]{navarro97}  
%%%%%%%%%%
\begin{equation}
u(r|M)=\frac{\rho(r)}{M}=\frac{\rho_s}{M(r/r_s)(1+r/r_s)^2},
\end{equation}
%%%%%%%%%%
and is truncated at $r_{\rm 200c}$.
An important parameter to characterize the NFW profile is
concentration $c_{\rm 200c}=r_{\rm 200c}/r_s$. We use the 
mass-concentration relation from \citet{duffy08}
%%%%%%%%%%
\begin{equation}
c_{\rm 200c}=\frac{5.71}{(1+z)^{0.47}}\left(\frac{M}
{2\times 10^{12}h^{-1}M_\odot}\right)^{-0.084}.
\end{equation}
%%%%%%%%%%
Similarly, the 1-halo term of the average number of neighboring
galaxies around a satellite galaxy at distance $r$ from the
halo center is derived as
%%%%%%%%%%
\begin{eqnarray}
\langle N_{\rm p}^{{\rm 1h,s}(r)}\rangle_M
&=&\int_0^\infty dk\, r_{\rm p,max}J_1(kr_{\rm p,max})\nonumber\\
&&\hspace*{-8mm}\times
\left[\langle N_{\rm cen}(M)\rangle +\langle N_{\rm sat}(M)
\rangle \tilde{u}(k|M)\right]j_0(kr).
\end{eqnarray}
%%%%%%%%%%
In each halo with mass $M$, we can also define the probability of a
member galaxy being a central or a satellite galaxy at $r$ as
%%%%%%%%%%
\begin{equation}
p({\rm c}|M)=\frac{\langle N_{\rm cen}(M)\rangle}{\langle N(M)\rangle},
\end{equation}
%%%%%%%%%%
\begin{equation}
p({\rm s},r|M)=\frac{\langle N_{\rm sat}(M)\rangle}{\langle
  N(M)\rangle} u(r|M).
\end{equation}
%%%%%%%%%%
The average neighboring galaxy counts within the same halo are then
computed as
%%%%%%%%%%
\begin{eqnarray}
\langle N_{\rm p}^{\rm 1h}\rangle_M
&=&\langle N_{\rm p}^{\rm 1h,c}\rangle_M p({\rm c}|M)\nonumber\\
&&+\int_0^{r_{\rm 200c}} 
4\pi r^2dr \langle N_{\rm p}^{{\rm 1h,s}(r)}\rangle_M
p({\rm s},r|M),\nonumber\\
&=&\int_0^\infty dk\, r_{\rm p,max}J_1(kr_{\rm p,max})
\frac{1}{\langle N(M)\rangle}\nonumber\\
&&\hspace*{-21mm}\times
\left[2\langle N_{\rm cen}(M)\rangle\langle N_{\rm sat}(M)\rangle
\tilde{u}(k|M)+\langle N_{\rm sat}(M)\rangle^2
\tilde{u}(k|M)^2\right].
\label{eq:ngc_ave_1h}
\end{eqnarray}
%%%%%%%%%%
On the other hand, the probability distribution of the host dark halo
mass is given by
%%%%%%%%%%
\begin{equation}
p(M)=\frac{1}{\bar{n}}\langle N(M)\rangle\frac{dn}{dM},
\label{eq:pm}
\end{equation}
%%%%%%%%%%
where $dn/dM$ is the mass function of dark matter halos, for which we adopt
the fitting function of \citet{tinker08}, and $\bar{n}$ is
the average number density of the target/neighboring galaxy population
defined by  
%%%%%%%%%%
\begin{equation}
\bar{n}=\int_0^\infty dM\langle N(M)\rangle\frac{dn}{dM}.
\label{eq:n3dave}
\end{equation}
%%%%%%%%%%

Hence the average neighboring galaxy counts from galaxies residing in
halos of all masses reduce to  
%%%%%%%%%%
\begin{eqnarray}
\langle N_{\rm p}^{\rm 1h}\rangle &=&
\int_0^\infty dM\langle N_{\rm p}^{\rm 1h}\rangle_M p(M)\nonumber\\
&=& \bar{n} \int_0^\infty dk \, r_{\rm  p,max}
P^{\rm 1h}(k) J_1(kr_{\rm p,max})\nonumber \\
&= & \bar{n} \int dV_{\rm c} \xi^{\rm 1h}(\mathbf{x}),
\label{eq:np_1h_all}
\end{eqnarray}
%%%%%%%%%%
where $P^{\rm 1h}(k)$ is the standard 1-halo term of galaxy power
spectrum 
%%%%%%%%%%
\begin{eqnarray}
P^{\rm 1h}(k)&=&\frac{1}{\bar{n}^2}\int_0^\infty 
dM \frac{dn}{dM} \nonumber\\
&& \hspace*{-21mm}\times \left[2\langle N_{\rm
    cen}(M)\rangle\langle N_{\rm sat}(M)\rangle 
\tilde{u}(k|M)+\langle N_{\rm sat}(M)\rangle^2\tilde{u}(k|M)^2\right].
\end{eqnarray}
%%%%%%%%%%

%%%%%%%%%%%%%%%%%%%%%%%%%%%%%%%%%%%%%%%%%%%%%%%%%%%%%%%%%%%%%%%%%%%%%%
\subsubsection{2-halo term and background}
%%%%%%%%%%%%%%%%%%%%%%%%%%%%%%%%%%%%%%%%%%%%%%%%%%%%%%%%%%%%%%%%%%%%%%

Next we consider the 2-halo term, i.e., contributions to neighboring
galaxy counts from different halos. We shall include contributions
from both the spatially correlated halos and uncorrelated structures
along the line-of-sight. 

We start by describing neighboring galaxy counts formally as
%%%%%%%%%%
\begin{equation}
N_{\rm p}(<r_{\rm p, max})=\int dV' n_{\rm t}(\mathbf{x}')
\int dV_{\rm c} n_{\rm n}(\mathbf{x}+\mathbf{x}'),
\label{eq:np_formal}
\end{equation}
%%%%%%%%%%
where $n_{\rm t}(\mathbf{x})$ is the number density of the target
galaxy. The number density of neighboring galaxies is written using
the density fluctuation $\delta_{\rm n}=\delta_{\rm n}(\mathbf{x})$ as 
$n_{\rm n}(\mathbf{x})=\bar{n}(1+\delta_{\rm n})$.  
Since we are interested in a single
target galaxy, $n_{\rm t}(\mathbf{x})$ is written as 
%%%%%%%%%%
\begin{equation}
n_{\rm t}(\mathbf{x})=\bar{n}_{\rm t}(1+\delta_{\rm t})=\delta_{\rm D}
(\mathbf{x}-\mathbf{x}_{\rm t}),
\end{equation}
%%%%%%%%%%
where $\delta_{\rm D}(\mathbf{x})$ denotes the Dirac delta function, 
$\mathbf{x}_{\rm t}$ is the location of that galaxy, and
%%%%%%%%%%
\begin{equation}
\int dV n_{\rm t}(\mathbf{x})=\int dV \bar{n}_{\rm t}=1,
\end{equation}
%%%%%%%%%%
where the integral is performed over a sufficiently large volume.

The average 2-halo neighboring galaxy counts for the target galaxy in
a halo of mass $M$, including the background contribution, are
obtained from Equation~(\ref{eq:np_formal}) as 
%%%%%%%%%%
\begin{equation}
\langle N_{\rm p}^{\rm 2h}\rangle_M=\bar{N}_{\rm n}
+\Delta^{\rm2h}_{N,{\rm tn}}.
\end{equation}
%%%%%%%%%%
The first term describes the background contribution (or random chance
projection), while the second term describes the enhancement of
neighboring galaxy counts due to clustering of the underlying matter
density field. Each term is explicitly written as
%%%%%%%%%%
\begin{equation}
\bar{N}_{\rm n}=\bar{n}\int dV_{\rm c},
\end{equation}
%%%%%%%%%%
\begin{eqnarray}
\Delta^{\rm 2h}_{N,{\rm tn}}&=&
\bar{n}\int dV_{\rm c} \xi^{\rm 2h}_{\rm tn}(\mathbf{x})\nonumber\\
&= &
\bar{n}
\int_0^\infty dk\,r_{\rm p,max}P^{\rm 2h}_{\rm tn}(k)J_1(kr_{\rm p,max}). 
\end{eqnarray}
%%%%%%%%%%
In practice, there are additional terms originating from higher order
correlation functions \citep[see Sec. 39 of][]{peebles80}. In this
paper we ignore those higher order contributions for simplicity. As we
will see below, this is a reasonably good approximation in our case
where we study number counts projected over redshift ranges that is
much larger than typical correlating lengths ($\mathcal{O}(10)$~Mpc).
The 2-halo term of the power spectrum is simply given by
%%%%%%%%%%
\begin{equation}
P^{\rm 2h}_{\rm tn}(k)=b(M)\bar{b}P_m(k),
\end{equation}
%%%%%%%%%%
where $b(M)$ is the halo bias for which we adopt a fitting function of
\citet{tinker10}, $P_m(k)$ is the linear matter power spectrum
(computed using the fitting formulae of \citealt{eisenstein99}), 
and $\bar{b}$ is the average galaxy bias computed by 
%%%%%%%%%%
\begin{equation}
\bar{b}=\frac{1}{\bar{n}}
\int_0^\infty  dM\,b(M)\langle N(M)\rangle\frac{dn}{dM}.
\end{equation}
%%%%%%%%%%
Again by averaging $\langle N_{\rm p}^{\rm 2h}\rangle_M$ over $p(M)$
(Equation~\ref{eq:pm}) we deduce an expression similar to
Equation~(\ref{eq:np_1h_all}) 
%%%%%%%%%%
\begin{eqnarray}
\langle N_{\rm p}^{\rm 2h}\rangle &=&
\int_0^\infty dM\langle N_{\rm p}^{\rm 2h}\rangle_M p(M)\nonumber\\
&=& \bar{N}_{\rm n}+
\bar{n} \int dV_{\rm c} \xi^{\rm 2h}(\mathbf{x}),
\label{eq:ngc_ave_2h}
\end{eqnarray}
%%%%%%%%%%
where $\xi^{\rm 2h}(\mathbf{x})$ is the Fourier transform of the
two-halo galaxy power spectrum $P^{\rm 2h}(k)=\bar{b}^2P_m(k)$.

Based on this formalism we can also derive the variance of neighboring
galaxy counts in the limit of $\int dV\rightarrow\infty$ as
%%%%%%%%%%
\begin{equation}
(\sigma^{\rm 2h}_{\rm p})^2\equiv
\langle N_{\rm p}^{\rm 2h}N_{\rm p}^{\rm 2h}\rangle_M
-\langle N_{\rm p}^{\rm 2h}\rangle_M^2 =
\bar{N}_{\rm n}+\bar{N}_{\rm n}\Delta^{\rm 2h}_{N,{\rm nn}},
\label{eq:ngc_var_2h}
\end{equation}
%%%%%%%%%%
where 
%%%%%%%%%%
\begin{eqnarray}
\Delta^{\rm 2h}_{N,{\rm nn}}&=&\frac{\bar{n}^2}{\bar{N}_{\rm n}}
\int dV_{\rm c} \int dV_{\rm c}'\xi_{\rm nn}^{\rm
  2h}(\mathbf{x}-\mathbf{x}')\nonumber\\
&= &\bar{n}\int_0^\infty \frac{2dk}{k}P^{\rm 2h}_{\rm nn}(k)
\left[J_1(kr_{\rm p,max})\right]^2,
\end{eqnarray}
%%%%%%%%%%
with $P^{\rm 2h}_{\rm nn}(k)=P^{\rm 2h}(k)=\bar{b}^2P_m(k)$.
Again, we have ignored contributions from higher order correlations. 

%%%%%%%%%%%%%%%%%%%%%%%%%%%%%%%%%%%%%%%%%%%%%%%%%%%%%%%%%%%%%%%%%%%%%%
\section{Probability distributions of neighboring galaxy counts
  and host dark halo masses}
\label{sec:pdf}
%%%%%%%%%%%%%%%%%%%%%%%%%%%%%%%%%%%%%%%%%%%%%%%%%%%%%%%%%%%%%%%%%%%%%%

%%%%%%%%%%%%%%%%%%%%%%%%%%%%%%%%%%%%%%%%%%%%%%%%%%%%%%%%%%%%%%%%%%%%%%
\subsection{Probability distribution of neighboring galaxy counts}
%%%%%%%%%%%%%%%%%%%%%%%%%%%%%%%%%%%%%%%%%%%%%%%%%%%%%%%%%%%%%%%%%%%%%%

We start by deriving the PDF of neighboring galaxy counts around a
target galaxy that resides in a dark halo with mass $M$. For this
purpose, we adopt the following assumptions: (1) the PDF of 1-halo
neighboring galaxy counts follows the Poisson distribution, which is a
reasonable assumption at least for massive halos where $\langle
N(M)\rangle$ is large \citep[e.g.,][]{lin04,zheng05}, (2) the PDF of
2-halo plus background neighboring galaxy counts follows the
log-normal distribution, given that the probability distribution
function of cosmological density fields is accurately described by the
log-normal distribution \citep[e.g.,][]{kayo01}, and (3) the 1-halo
and 2-halo PDFs are uncorrelated with each other.  

For the expected value of $\lambda$, the Poisson distribution gives
the PDF of observed number $N$ as
%%%%%%%%%%
\begin{equation}
p^{\rm P}(N|\lambda)
=\frac{\lambda^N e^{-\lambda}}{N!}.
\end{equation}
%%%%%%%%%%
We include the position dependence of the galaxy in the calculation of
the PDF of 1-halo neighboring galaxy counts. Using the results given
in Section~\ref{sec:1h}, we obtain
%%%%%%%%%%
\begin{eqnarray}
p^{\rm 1h}(N_{\rm p}^{\rm 1h}|M)&=&
p^{\rm P}(N_{\rm p}^{\rm 1h}|\langle N_{\rm p}^{\rm 1h, c}\rangle_M)
p({\rm c}|M)\nonumber\\
&&\hspace*{-10mm}+\int_0^{r_{\rm 200c}}4\pi r^2dr \,p^{\rm P}
(N_{\rm p}^{\rm 1h}|\langle N_{\rm p}^{{\rm s}(r)}\rangle_M)
p({\rm s},r|M).
\label{eq:pdf_ngc_1h}
\end{eqnarray}
%%%%%%%%%%
We assume that the PDF of 2-halo neighboring galaxy counts follows the
log-normal distribution
%%%%%%%%%%
\begin{equation}
p^{\rm LN}(N|\mu, s)
=\frac{1}{N\sqrt{2\pi}s}e^{-\frac{(\ln N-\mu)^2}{2s^2}}.
\label{eq:logn}
\end{equation}
%%%%%%%%%%
Detailed discussions on the validity of this assumption are given in Appendix.
We then obtain
%%%%%%%%%%
\begin{equation}
p^{\rm 2h}(N_{\rm p}^{\rm 2h}|M)=p^{\rm LN}(N_{\rm p}^{\rm 2h}|\mu_M, s_M),
\end{equation}
%%%%%%%%%%
where the parameters $\mu_M$ and $s_M$ are related to the mean 
$\langle N_{\rm p}^{\rm 2h}\rangle$ and variance 
$(\sigma^{\rm 2h}_{\rm p})^2$ as
%%%%%%%%%%
\begin{equation}
\mu_M=\ln\frac{\langle N_{\rm p}^{\rm 2h}\rangle^2}
{\sqrt{\langle N_{\rm p}^{\rm 2h}\rangle^2+(\sigma^{\rm 2h}_{\rm p})^2}},
\end{equation}
%%%%%%%%%%
\begin{equation}
s^2_M=\ln\left[1+\frac{(\sigma^{\rm 2h}_{\rm p})^2}
{\langle N_{\rm p}^{\rm 2h}\rangle^2}\right].
\end{equation}
%%%%%%%%%%
Finally we combine these two PDFs to obtain the PDF of total neighboring
galaxy counts 
%%%%%%%%%%
\begin{equation}
p(N_{\rm p}|M)=\sum_{N=0}^{N_{\rm p}} p^{\rm 1h}(N_{\rm p}-N|M)
p^{\rm 2h}(N|M).
\label{eq:pofnp}
\end{equation}
%%%%%%%%%%

%%%%%%%%%%%%%%%%%%%%%%%%%%%%%%%%%%%%%%%%%%%%%%%%%%%%%%%%%%%%%%%%%%%%%%
\subsection{Probability distribution of host halo masses}
%%%%%%%%%%%%%%%%%%%%%%%%%%%%%%%%%%%%%%%%%%%%%%%%%%%%%%%%%%%%%%%%%%%%%%
We now turn the problem around and ask how well we can infer the host
halo mass of a single target galaxy from observed neighboring galaxy
counts. This is obtained by using the Bayes' theorem
%%%%%%%%%%
\begin{equation}
p(M|N_{\rm p})=\frac{p(N_{\rm p}|M)p(M)}{\int dM p(N_{\rm p}|M)p(M)},
\label{eq:pofm}
\end{equation}
%%%%%%%%%%
where $p(N_{\rm p}|M)$ and $p(M)$ are given in
Equations~(\ref{eq:pofnp}) and (\ref{eq:pm}), respectively. 

%%%%%%%%%%%%%%%%%%%%%%%%%%%%%%%%%%%%%%%%%%%%%%%%%%%%%%%%%%%%%%%%%%%%%%
\section{Comparison with the mock galaxy catalog}
\label{sec:mock}
%%%%%%%%%%%%%%%%%%%%%%%%%%%%%%%%%%%%%%%%%%%%%%%%%%%%%%%%%%%%%%%%%%%%%%

%%%%%%%%%%%%%%%%%%%%%%%%%%%%%%%%%%%%%%%%%%%%%%%%%%%%%%%%%%%%%%%%%%%%%%
\subsection{Mock galaxy catalog}
%%%%%%%%%%%%%%%%%%%%%%%%%%%%%%%%%%%%%%%%%%%%%%%%%%%%%%%%%%%%%%%%%%%%%%
We use the all-sky light-cone mock catalog from \citet{henriques12} to
test our analytic predictions.  This mock is based on the
semi-analytic model of \citet{guo11}, and is constructed by
replicating the Millennium Run simulation box ($500\,h^{-1}$Mpc on a
side) without transformations such as rotation, translation, or
inversion.  \citet{guo11} showed that their model can reproduce
important statistical properties of nearby galaxies such as the
stellar mass function and two-point correlation function. 
 
For each galaxy in the mock, information such as the sky position,
redshift, apparent and absolute magnitudes, host dark matter halo
mass, and central/satellite designation, is available. The catalog is
flux limited to $i<21$. In this paper, we generate a volume-limited
galaxy subsample of $M_i<-20$ to about $z=0.2$ and use the subsample
for our analysis. 

%%%%%%%%%%%%%%%%%%%%%%%%%%%%%%%%%%%%%%%%%%%%%%%%%%%%%%%%%%%%%%%%%%%%
\begin{figure}
\epsscale{0.95}
\plotone{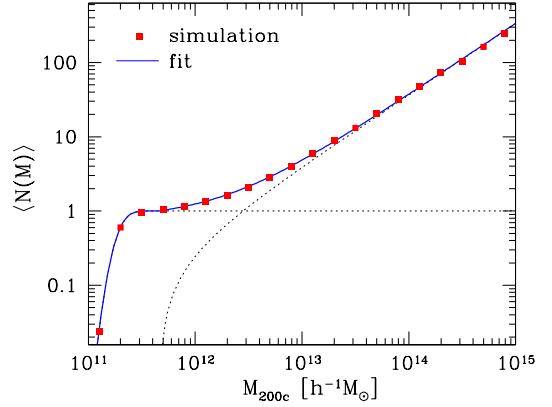}
%\vspace{-4mm}
\caption{ 
  The mean halo occupation number  of the mock galaxy catalog. Filled
  squares show the mean occupation number directly obtained from the
  mock catalog. The solid line is best-fitting halo occupation number
  with Equations~(\ref{eq:hod_cen}) and (\ref{eq:hod_sat}). Dotted
  curves represent contributions from the central and satellite
  components. Best-fit parameters are summarized in
  Table~\ref{tab:hod}. 
  }
\label{fig:hod}
\end{figure}

%%%%%%%%%%%%%%%%%%%%%%%%%%%%%%%%%%%%%%%%%%%%%%%%%%%%%%%%%%%%%%%%%%%%

%%%%%%%%%%%%%%%%%%%%%%%%%%%%%%%%%%%%%%%%%%%%%%%%%%%%%%%%%%%%%%%%%%%%%
\begin{deluxetable}{cc}
\tablecaption{Best-fit HOD parameters}
%\tablewidth{0pt}
%
\tablehead{
\colhead{Parameter} & \colhead{Best-fit value} 
}
\startdata
  $M_{\rm min}[h^{-1}M_\odot]$ & $10^{11.277}$\\
  $\sigma$                  & 0.207\\
  $\kappa M_{\rm cut}[h^{-1}M_\odot]$ & $10^{11.666}$\\
  $M_1[h^{-1}M_\odot]$ & $10^{12.370}$\\
  $\alpha$ & 0.960
\enddata
\label{tab:hod}
\end{deluxetable}
%%%%%%%%%%%%%%%%%%%%%%%%%%%%%%%%%%%%%%%%%%%%%%%%%%%%%%%%%%%%%%%%%%%%%

%%%%%%%%%%%%%%%%%%%%%%%%%%%%%%%%%%%%%%%%%%%%%%%%%%%%%%%%%%%%%%%%%%%%%

%%%%%%%%%%%%%%%%%%%%%%%%%%%%%%%%%%%%%%%%%%%%%%%%%%%%%%%%%%%%%%%%%%%%%%
\subsection{Halo occupation number}
%%%%%%%%%%%%%%%%%%%%%%%%%%%%%%%%%%%%%%%%%%%%%%%%%%%%%%%%%%%%%%%%%%%%%%
An essential ingredient in our formalism is the halo occupation number
$\langle N(M)\rangle$. In the mock galaxy catalog, we know host halo
properties of individual galaxies, which allows us to derive the mean
halo occupation number directly without ambiguity. We fit the halo
occupation number obtained directly from the mock catalog to the
parametrized form in Equations~(\ref{eq:hod_cen}) and
(\ref{eq:hod_sat}), which contains five free parameters in total. We
show the fitting result in Figure~\ref{fig:hod}, and summarize
best-fit parameters in Table~\ref{tab:hod}. We find that the HOD model
adopted in this paper reproduces the mean occupation number in the
mock galaxy catalog very well.   

%%%%%%%%%%%%%%%%%%%%%%%%%%%%%%%%%%%%%%%%%%%%%%%%%%%%%%%%%%%%%%%%%%%%%%
\subsection{Neighboring galaxy counts}
%%%%%%%%%%%%%%%%%%%%%%%%%%%%%%%%%%%%%%%%%%%%%%%%%%%%%%%%%%%%%%%%%%%%%%

%%%%%%%%%%%%%%%%%%%%%%%%%%%%%%%%%%%%%%%%%%%%%%%%%%%%%%%%%%%%%%%%%%%%
\begin{figure}
\epsscale{0.95}
\plotone{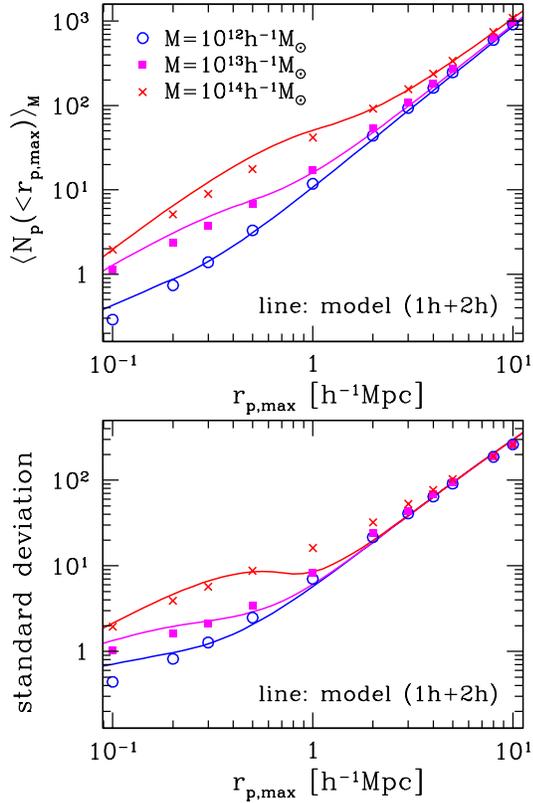}
%\vspace{-4mm}
\caption{ 
  The average neighboring galaxy counts within the comoving
  transverse distance $r_{\rm p,max}$. The number counts are computed
  around galaxies at $z=0.15$ which reside in host halos with mass
  $M$. The redshift range of the neighboring galaxies is set to
  $z=0.15\pm 0.02$. Symbols show the result from the mock galaxy
  catalog for host halo masses of $M=10^{12}$ ({\it open circles}),
  $10^{13}$ ({\it filled squares}), and $10^{14}h^{-1}M_\odot$ ({\it
    crosses}), and solid lines show corresponding analytic model
  predictions including both 1-halo and 2-halo contributions. The
  upper panel shows the average neighboring galaxy counts 
   $\langle N_{\rm p}(<r_{\rm p,max})\rangle_M$ 
  (Equation~\ref{eq:ngc_ave}), whereas the lower panel shows its 
   standard deviation $\sigma_{\rm p}$ (Equation~\ref{eq:ngc_var}).  
}
\label{fig:nc}
\end{figure}

%%%%%%%%%%%%%%%%%%%%%%%%%%%%%%%%%%%%%%%%%%%%%%%%%%%%%%%%%%%%%%%%%%%%

Before comparing the PDF of the halo mass, we first check whether our
analytic model reproduces neighboring galaxy counts in the mock galaxy
catalog. The average (projected) neighboring galaxy counts around a
galaxy inside a halo with mass $M$ is given by 
%%%%%%%%%%
\begin{equation}
\langle N_{\rm p}(<r_{\rm p,max})\rangle_M =
\langle N_{\rm p}^{\rm 1h}\rangle_M+\langle N_{\rm p}^{\rm
  2h}\rangle_M,
\label{eq:ngc_ave}
\end{equation}
%%%%%%%%%%
where $\langle N_{\rm p}^{\rm 1h}\rangle_M$ and $\langle N_{\rm p}^{\rm
  2h}\rangle_M$ are given in Equations~(\ref{eq:ngc_ave_1h}) and
(\ref{eq:ngc_ave_2h}), respectively.

We compute neighboring galaxy counts as a function of the maximum
comoving transverse distance $r_{\rm p,max}$. The redshift of target
galaxies for which neighboring galaxy counts are considered is
arbitrarily fixed at $z=0.15$. In the mock galaxy catalog, we actually
use galaxies in the small redshift range of $z=0.15\pm
0.005$. Throughout the paper we consider the redshift range of
neighboring galaxies of $\Delta z=0.02$, which is sufficiently large
to include the correlated structure.  

Figure~\ref{fig:nc} (upper panel) compares the average neighboring
galaxy counts from the mock galaxy catalog with the analytic
calculation (Equation~\ref{eq:ngc_ave}). We consider target galaxies
living in halos of three masses, $M=10^{12}$, $10^{13}$, and 
$10^{14}h^{-1}M_\odot$, and find that our model agrees well with the
mock result, both at small and large $r_{\rm p,max}$ where 1-halo and
2-halo contributions are dominated, respectively. As expected,
neighboring galaxy counts are higher for galaxies in more massive host
halos. 

We also check the variance of the average neighboring galaxy counts,
$\sigma_{\rm p}^2\equiv \langle N_{\rm p}N_{\rm p}\rangle_M
-\langle N_{\rm p}\rangle_M^2$. Again, the variance is given by the
sum of 1-halo and 2-halo contributions
%%%%%%%%%%
\begin{equation}
\sigma_{\rm p}^2=(\sigma_{\rm p}^{\rm 1h})^2+
(\sigma_{\rm p}^{\rm 2h})^2,
\label{eq:ngc_var}
\end{equation}
%%%%%%%%%%
where $(\sigma_{\rm p}^{\rm 2h})^2$ is derived in
Equation~(\ref{eq:ngc_var_2h}). Given the PDF of 1-halo neighboring
galaxy counts (Equation~\ref{eq:pdf_ngc_1h}), we can compute 
$(\sigma_{\rm p}^{\rm 1h})^2$ as
%%%%%%%%%%
\begin{eqnarray}
(\sigma_{\rm p}^{\rm 1h})^2&=&
\langle N_{\rm p}^{\rm 1h}\rangle_M-\langle N_{\rm p}^{\rm
  1h}\rangle_M^2+\langle N_{\rm p}^{\rm
  1h,c}\rangle_M^2 p({\rm c}|M)
\nonumber\\&&
+\int_0^{r_{\rm 200c}}4\pi r^2dr \langle N_{\rm p}^{{\rm 1h,s}(r)}\rangle_M^2
p({\rm s},r|M).
\end{eqnarray}
%%%%%%%%%%
The comparison shown in Figure~\ref{fig:nc} (lower panel) indicates
that the analytic model reproduces the variance as well, in both
1-halo and 2-halo regimes. However there is relatively large deviation
of the analytic model at $r_{\rm p,max}\sim 1h^{-1}{\rm Mpc}$ for the
host halo mass of $M=10^{14}h^{-1}M_\odot$. This might be explained by
the non-Gaussian error of the 2-halo term at very small scale. 
Another possible explanation of the discrepancy is the presence of
satellite galaxies outside $r_{\rm 200c}$; we find that the mock
galaxy catalog contains satellite galaxies that are located outside
$r_{\rm 200c}$ ($\sim 10$~\%), whereas we have assumed that all
satellite galaxies are located within $r_{\rm 200c}$. 

%%%%%%%%%%%%%%%%%%%%%%%%%%%%%%%%%%%%%%%%%%%%%%%%%%%%%%%%%%%%%%%%%%%%%%
\subsection{PDF of host halo masses}
%%%%%%%%%%%%%%%%%%%%%%%%%%%%%%%%%%%%%%%%%%%%%%%%%%%%%%%%%%%%%%%%%%%%%%

\begin{figure}
\epsscale{1}
\plotone{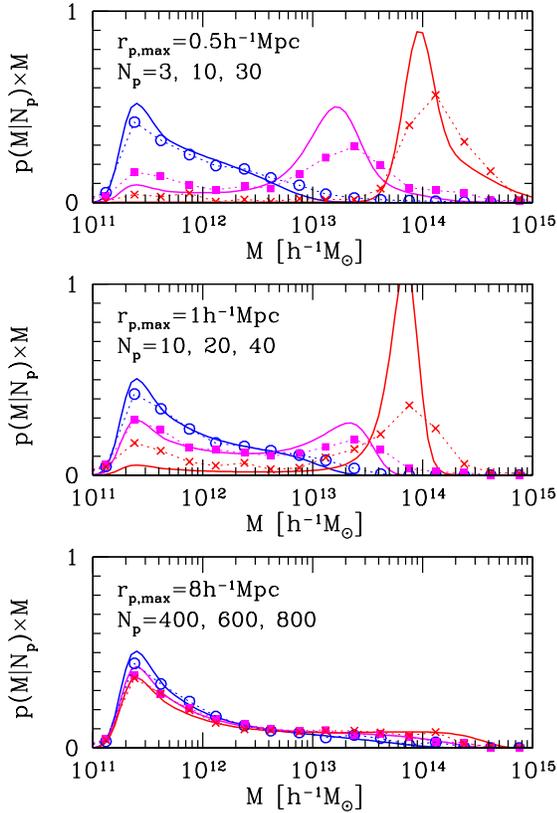}
%\vspace{-4mm}
\caption{ 
  PDFs of the host halo masses of galaxies with neighboring
  galaxy counts within $r_{\rm p,max}$ being $N_{\rm p}$, $p(M|N_{\rm
    p})$. From top to bottom panels, we consider three different
  maximum comoving transverse distance, $r_{\rm p,max}=0.5$, $1$, and
  $8h^{-1}{\rm Mpc}$, within which the number of neighboring galaxy 
  is counted. Different symbols show conditional PDFs for different
  neighboring galaxy counts $N_{\rm p}$ from the mock galaxy catalog, 
  and corresponding analytic model predictions (Equation~\ref{eq:pofm}) 
  are plotted by solid lines. 
  }
\label{fig:pofm}
\end{figure}

We now compare our model of PDFs of the host halo masses given
neighboring galaxy counts, $p(M|N_{\rm p})$ (Equation~\ref{eq:pofm}),
with those obtained directly from the mock galaxy catalog. 
Figure~\ref{fig:pofm} compares the PDFs from the mock galaxy catalog
with our analytic model. We find that our analytic models are in
reasonably good agreement with the simulation result.  In particular
the analytic model nicely captures the complex behaviors of the PDFs.

%%%%%%%%%%%%%%%%%%%%%%%%%%%%%%%%%%%%%%%%%%%%%%%%%%%%%%%%%%%%%%%%%%%%
%%%%%%%%%%%%%%%%%%%%%%%%%%%%%%%%%%%%%%%%%%%%%%%%%%%%%%%%%%%%%%%%%%%%

We find the PDF exhibits two peaks when the number of neighboring
galaxies $N_{\rm p}$ is large. The higher mass peak corresponds to the
case that the galaxy resides in massive halos with about the peak
mass. On the other hand, the lower mass peak is due to the chance
projection, i.e., the low-mass host halo of the galaxy is superposed
on a massive halo or a dense structure on the sky, which boosts the
neighboring galaxy counts. Our analytic model reproduces this behavior.

We find that the dependence of the PDFs on neighboring galaxy counts is
weak for the large cylinder radius of $r_{\rm p,max}=8 h^{-1}{\rm
  Mpc}$ for which the contribution is dominated by the 2-halo term. Although
neighboring galaxy counts depend on the host halo mass due to the halo
mass dependence of the halo bias, neighboring galaxy counts also involves
the large scatter  originating from clustering of neighboring galaxies
(see Figure~\ref{fig:nc}). Thus, in order to infer the host halo mass
of individual galaxies, small cylinder radii of $r_{\rm p,max}\la 1
h^{-1}{\rm Mpc}$ should be used for neighboring galaxy counts. 

%%%%%%%%%%%%%%%%%%%%%%%%%%%%%%%%%%%%%%%%%%%%%%%%%%%%%%%%%%%%%%%%%%%%
\begin{figure}
\epsscale{0.9}
\plotone{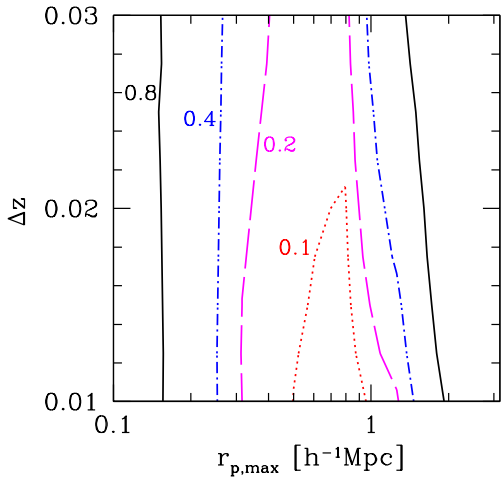}
%\vspace{-4mm}
\caption{ 
  Contours of the variance of inferred host halo masses,
  $\sigma^2_{\log M}$ which is computed from the conditional PDF
  $p(M|N_{\rm p})$, in the $r_{\rm p,max}$-$\Delta z$ plane. The
  input halo mass is $M=10^{14}h^{-1}M_\odot$, and the average
  neighboring number $\langle N_{\rm p}(<r_{\rm p,max})\rangle_M$
  (Equation~\ref{eq:ngc_ave}) is used for the input $N_{\rm p}$ to
  compute the PDF. Dotted, dashed, dot-dashed, and solid lines show
  contours of $\sigma^2_{\log M}=0.1$, $0.2$, $0.4$, and $0.8$,
  respectively. 
  }
\label{fig:wpm}
\end{figure}
%%%%%%%%%%%%%%%%%%%%%%%%%%%%%%%%%%%%%%%%%%%%%%%%%%%%%%%%%%%%%%%%%%%%

With the analytic model of PDFs of the host halo masses, we can easily
explore how different choices of neighboring counts provide tighter
constraints on halo masses. In addition to $r_{\rm p,max}$ discussed
above, we can also adjust the redshift range $\Delta z$. Here we
demonstrate how we can 
optimize $r_{\rm p,max}$ and $\Delta z$ from the PDFs. As a specific
example, we consider an input halo mass of $M=10^{14}h^{-1}M_\odot$. 
We first compute the average neighboring number $\langle N_{\rm
  p}(<r_{\rm p,max})\rangle_M$ (Equation~\ref{eq:ngc_ave}) for given
values of $r_{\rm p,max}$ and $\Delta z$, and compute the conditional
PDF $p(M|N_{\rm p})$ (Equation~\ref{eq:pofm}) given the number count
$N{\rm p}$ equal to the average neighboring number. We quantify the
tightness of the halo mass inference by computing the variance of the
halo mass $\log M$, $\sigma^2_{\log M}$, using the computed
PDF. Smaller $\sigma^2_{\log M}$ indicates a narrower peak of the
PDF and therefore tighter constraints on the halo mass. The contours
shown in Figure~\ref{fig:wpm} confirm that cylinder radii of $r_{\rm
  p,max}\sim 0.5-1 h^{-1}{\rm Mpc}$ provide tightest constraints on
the halo masses. We find that the dependence on $\Delta z$ is not very
strong, but smaller $\Delta z$ generally produces tighter constraints,
because of the smaller number of projected galaxies. We note that the
best choice of the parameters depends also on other factors such as
halo masses and underlying HOD models.

%%%%%%%%%%%%%%%%%%%%%%%%%%%%%%%%%%%%%%%%%%%%%%%%%%%%%%%%%%%%%%%%%%%%%%
\section{Extensions and applications}
\label{sec:ext}
%%%%%%%%%%%%%%%%%%%%%%%%%%%%%%%%%%%%%%%%%%%%%%%%%%%%%%%%%%%%%%%%%%%%%%

In Section~\ref{sec:mock} we have used a mock galaxy catalog to
demonstrate that our analytic model can robustly predict the PDF of
halo mass for a {\it single} galaxy, given its neighboring counts.
Although the example we have considered is somewhat idealized (i.e.,
with 100\% spectroscopic sampling and redshift success rate), we note
that these conditions are not required.  In principle, for a given
galaxy sample, as long as we can accurately model its HOD, our model
can be used to predict the host halo mass PDF for the galaxies {\it
  in} the sample. 

Here we briefly discuss various ways to extend the scope and
capabilities of our model, including the possibility to distinguish
central galaxies from satellites, the situation when the target and
neighboring galaxies are from two different galaxy samples, the
potential of improving the PDF with neighboring counts within multiple
apertures, constraints on the HOD itself, the application to galaxy
samples selected with photometric redshifts, and the prospect of
applying the same principle to inferring masses of galaxy clusters. 

%%%%%%%%%%%%%%%%%%%%%%%%%%%%%%%%%%%%%%%%%%%%%%%%%%%%%%%%%%%%%%%%%%%%%%
\subsection{Probability of being central}
%%%%%%%%%%%%%%%%%%%%%%%%%%%%%%%%%%%%%%%%%%%%%%%%%%%%%%%%%%%%%%%%%%%%%%

In many of HOD studies, central and satellite components are treated
separately. Therefore it is of great interest to know whether each
galaxy corresponds to the central or satellite type. This can be
obtained easily in this theoretical framework. From
Equation~(\ref{eq:pdf_ngc_1h}), we derive the joint probability of
being a central galaxy and having a 1-halo neighboring galaxy count of 
$N_{\rm p}^{\rm 1h}$ as
%%%%%%%%%%
\begin{eqnarray}
p^{\rm 1h}({\rm c},N_{\rm p}^{\rm 1h}|M)=
p^{\rm P}(N_{\rm p}^{\rm 1h}|\langle N_{\rm p}^{\rm 1h, c}\rangle_M)
p({\rm c}|M),
\end{eqnarray}
%%%%%%%%%%
By replacing $p^{\rm 1h}(N_{\rm p}-N|M)$ in Equation~(\ref{eq:pofnp})
with $p^{\rm 1h}({\rm c},N_{\rm p}-N|M)$ and inserting the result to
Equation~(\ref{eq:pofm}), we obtain the joint PDF of $p({\rm c},M|N_{\rm p})$. The  probability of being central given the host halo mass $M$
and neighboring galaxy count $N_{\rm p}$ is then obtained as
%%%%%%%%%%
\begin{eqnarray}
p({\rm c}|M, N_{\rm p})=\frac{p({\rm c},M|N_{\rm p})}{p(M|N_{\rm p})}.
\label{eq:pcen}
\end{eqnarray}
%%%%%%%%%%

%%%%%%%%%%%%%%%%%%%%%%%%%%%%%%%%%%%%%%%%%%%%%%%%%%%%%%%%%%%%%%%%%%%%
\begin{figure}
\epsscale{0.95}
\plotone{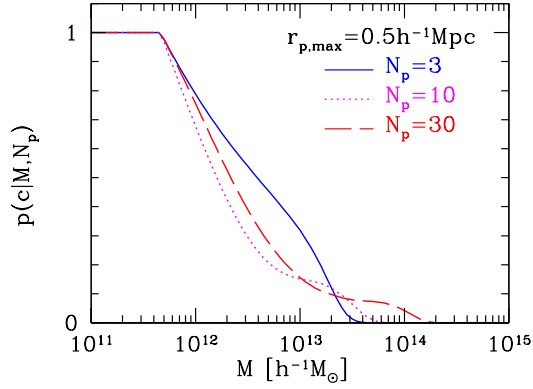}
%\vspace{-4mm}
\caption{ 
  The probability of being a central galaxy given the halo mass and
  neighboring galaxy counts (Equation~\ref{eq:pcen}). The cylinder
  radius is fixed to $r_{\rm p,max}=0.5h^{-1}{\rm Mpc}$. We consider
  three different $N_{\rm p}$: $N_{\rm p}=3$ ({\it solid}), $10$ 
  ({\it  dotted}), and $30$ ({\it dashed}).
    }
\label{fig:cent}
\end{figure}

%%%%%%%%%%%%%%%%%%%%%%%%%%%%%%%%%%%%%%%%%%%%%%%%%%%%%%%%%%%%%%%%%%%%

Figure~\ref{fig:cent} shows some examples of $p({\rm c}|M, N_{\rm p})$
for the cylinder of radius $r_{\rm p,max}=0.5h^{-1}{\rm Mpc}$. The
probability is generally higher for smaller halo masses, simply
because there are fewer satellite galaxies in lower mass halos (see
Figure~\ref{fig:hod}). On the other hand, the probabilities shown in
Figure~\ref{fig:cent} exhibit dependences on neighboring galaxy
counts $N_{\rm p}$, suggesting that neighboring galaxy counts do provide 
information when determining whether a galaxy is central or not.

%%%%%%%%%%%%%%%%%%%%%%%%%%%%%%%%%%%%%%%%%%%%%%%%%%%%%%%%%%%%%%%%%%%%%%
\subsection{The use of different galaxy populations}
%%%%%%%%%%%%%%%%%%%%%%%%%%%%%%%%%%%%%%%%%%%%%%%%%%%%%%%%%%%%%%%%%%%%%%

While we have assumed that the target galaxy for which we infer the
host halo mass and neighboring galaxies are the same population and
they share the same HOD, in principle we can use populations with
different HODs for neighboring galaxy counts (e.g., red and blue
galaxies, galaxies of different luminosity/stellar mass, or
groups/clusters instead of galaxies). The extension of our theoretical
framework to this situation is straightforward. Suppose the halo
occupation number of the target galaxy $\langle N^{\rm t}(M)\rangle$
differs from that of neighboring galaxies $\langle N^{\rm n}(M)\rangle$, 
the 1-halo neighboring galaxy counts are modified as
%%%%%%%%%%
\begin{equation}
\langle N_{\rm p}^{\rm 1h,c}\rangle_M
=\int_0^\infty dk\, r_{\rm p,max}
J_1(kr_{\rm p,max})\langle N^{\rm n}_{\rm sat}(M)\rangle
\tilde{u}_{\rm n}(k|M),
\end{equation}
%%%%%%%%%%
\begin{eqnarray}
\langle N_{\rm p}^{{\rm 1h,s}(r)}\rangle_M
&=&\int_0^\infty dk\, r_{\rm p,max}J_1(kr_{\rm p,max})\nonumber\\
&&\hspace*{-8mm}\times
\left[\langle N^{\rm n}_{\rm cen}(M)\rangle +\langle N^{\rm n}_{\rm sat}(M)
\rangle \tilde{u}_{\rm n}(k|M)\right]j_0(kr),
\end{eqnarray}
%%%%%%%%%%
\begin{equation}
p({\rm c}|M)=\frac{\langle N^{\rm t}_{\rm cen}(M)\rangle}{\langle N^{\rm
    t}(M)\rangle},
\end{equation}
%%%%%%%%%%
\begin{equation}
p({\rm s},r|M)=\frac{\langle N^{\rm t}_{\rm sat}(M)\rangle}{\langle
  N^{\rm t}(M)\rangle} u_{\rm t}(r|M),
\end{equation}
%%%%%%%%%%
where $u_{\rm t}$ and $u_{\rm n}$ are spatial distributions of
satellite components of target and neighboring galaxies, respectively,
and the PDF of the host halo mass becomes
%%%%%%%%%%
\begin{equation}
p(M)=\frac{1}{\bar{n}_{\rm t}}\langle N^{\rm t}(M)\rangle\frac{dn}{dM},
\end{equation}
%%%%%%%%%%
and the average bias for the 2-halo calculation becomes
%%%%%%%%%%
\begin{equation}
\bar{b}_{\rm n}=\frac{1}{\bar{n}_{\rm n}}
\int_0^\infty  dM\,b(M)\langle N^{\rm n}(M)\rangle\frac{dn}{dM},
\end{equation}
%%%%%%%%%%
where $\bar{n}_{\rm t}$ and $\bar{n}_{\rm n}$ are average number
densities (see Equation~\ref{eq:n3dave}) of target and neighboring
galaxies, respectively. Also note that $\bar{n}$ in $\langle N_{\rm
  p}^{\rm 2h, c}\rangle_M$ should be interpreted as $\bar{n}_{\rm n}$.
Since different galaxy populations probe different halo masses, the
use of different galaxy populations for neighboring galaxy counts can
help improve the inference of the host halo mass. 

%%%%%%%%%%%%%%%%%%%%%%%%%%%%%%%%%%%%%%%%%%%%%%%%%%%%%%%%%%%%%%%%%%%%%%
\subsection{The use of multiple apertures}
%%%%%%%%%%%%%%%%%%%%%%%%%%%%%%%%%%%%%%%%%%%%%%%%%%%%%%%%%%%%%%%%%%%%%%

The inference of the host halo mass may be improved further by
considering multiple apertures, i.e., combining neighboring galaxy
counts with different cylinder radii. In this case, we need to take
account of the covariance of neighboring galaxy counts carefully. 
We leave the exploration of this possibility for future work.

%%%%%%%%%%%%%%%%%%%%%%%%%%%%%%%%%%%%%%%%%%%%%%%%%%%%%%%%%%%%%%%%%%%%%%
\subsection{Improving the analytic model}
%%%%%%%%%%%%%%%%%%%%%%%%%%%%%%%%%%%%%%%%%%%%%%%%%%%%%%%%%%%%%%%%%%%%%%

There is also room for the improvement of our analytic model. For
instance, attempts to improve the halo model have been made by
including the halo exclusion effect and non-linearity of the 2-halo
power spectrum \citep[e.g.,][]{tinker05}. The evaluation of
non-Gaussian error may be necessary to improve the estimate of the
variance of 2-halo neighboring galaxy counts, 
$(\sigma^{\rm 2h}_{\rm p})^2$ (see Figure~\ref{fig:nc}). We have also
ignored the complexity of the halo properties, such as the scatter in
the concentration parameter and the halo triaxiality. 
The HOD model itself is also subject to possible improvements,
  e.g., by including assembly bias \citep{wang13,zentner14,lacerna14}
  and relative velocities of central galaxies \citep{skiba11}.

%%%%%%%%%%%%%%%%%%%%%%%%%%%%%%%%%%%%%%%%%%%%%%%%%%%%%%%%%%%%%%%%%%%%%%
\subsection{Constraints on HOD with counts-in-cylinders}
%%%%%%%%%%%%%%%%%%%%%%%%%%%%%%%%%%%%%%%%%%%%%%%%%%%%%%%%%%%%%%%%%%%%%%

Neighboring galaxy counts depend on the underlying HOD model, and
therefore in principle they can be used to constrain the HOD
itself. For this application, it is crucial to estimate and remove the
chance projection. One can infer the effect of pure chance projection
by estimating the average number density, but the 2-halo contribution
is not homogeneous but clustered around the target galaxy. The
theoretical framework developed in this paper may be used to estimate
the 2-halo contribution and its variance to derive more rigorous
constraints on HOD with counts-in-cylinders.

%%%%%%%%%%%%%%%%%%%%%%%%%%%%%%%%%%%%%%%%%%%%%%%%%%%%%%%%%%%%%%%%%%%%%%
\subsection{Extension to galaxy samples with photometric redshifts}
%%%%%%%%%%%%%%%%%%%%%%%%%%%%%%%%%%%%%%%%%%%%%%%%%%%%%%%%%%%%%%%%%%%%%%

Although our formalism is readily applicable to several spectroscopic
surveys such as Sloan Digital Sky Survey \citep[SDSS;][]{york00},
Galaxy And Mass Assembly 
\citep[GAMA;][]{driver11}, PRIsm MUlti-object Survey
\citep[PRIMUS;][]{coil11}, VIMOS Public Extragalactic Redshift Survey
\citep[VIPERS;][]{guzzo14}, and eventually Dark Energy Spectroscopic
Instrument \citep[DESI;][]{levi13}, and Subaru Prime Focus 
Spectrograph \citep[PFS;][]{takada14},  
it is highly desirable to extend our model to incorporate samples
selected with photometric redshifts, so that we could fully exploit
the potential offered by large imaging surveys such as Subaru Hyper
Suprime-Cam \citep[HSC;][]{miyazaki12} survey, Dark Energy
  Survey\footnote{http://www.darkenergysurvey.org/}, and Large
  Synoptic Survey Telescope \citep[LSST;][]{lsst09}. 
We plan to address this extension in a future paper.

%%%%%%%%%%%%%%%%%%%%%%%%%%%%%%%%%%%%%%%%%%%%%%%%%%%%%%%%%%%%%%%%%%%%%%
\subsection{Inferring galaxy cluster masses using neighboring cluster counts}
%%%%%%%%%%%%%%%%%%%%%%%%%%%%%%%%%%%%%%%%%%%%%%%%%%%%%%%%%%%%%%%%%%%%%%

Galaxy clusters have long been regarded as a powerful cosmological
probe, as their abundance is sensitively dependent on parameters such
as $\Omega_m$ and $\sigma_8$.  Accurate knowledge of cluster mass is
required, however, before any useful cosmological constraints can be
deduced \citep{weinberg13}.  Our formalism could be straightforwardly
applied to galaxy clusters, especially samples detected in
optical/near-IR imaging surveys, for which it has typically been
difficult to deduce mass reliably for individual clusters.  With the
knowledge of PDF of cluster mass, it should be possible to obtain
tighter constraints from cluster abundance. 

%%%%%%%%%%%%%%%%%%%%%%%%%%%%%%%%%%%%%%%%%%%%%%%%%%%%%%%%%%%%%%%%%%%%%%
\section{Summary}
\label{sec:summary}
%%%%%%%%%%%%%%%%%%%%%%%%%%%%%%%%%%%%%%%%%%%%%%%%%%%%%%%%%%%%%%%%%%%%%%

In the highly successful HOD framework, the dark matter halo mass is
the sole factor controlling the galaxy formation process. It is thus
critically important to be able to estimate the halo mass of galaxies.
Traditionally, this is achieved by techniques such as strong
gravitational lensing \citep[e.g.,][]{bolton08}, galaxy-galaxy
lensing \citep[e.g.,][]{mandelbaum06}, satellite kinematics
\citep[e.g.,][]{more09}, and large scale clustering
\citep[e.g,.][]{zehavi11}.  Except for the 
first method, all the others work in a statistical sense, that is,
they cannot be applied to a single galaxy.  An alternative method is
developed by \citet{yang05}, where the concept of abundance matching
is employed to a large group catalog which includes groups with only
one member. Under the assumption of one-to-one correspondence between
halo mass and stellar mass/luminosity content of the group members, a
halo mass is {\it assigned} to each of the groups.  However, with this
approach, little or no information on the PDF of the halo mass is
provided. Furthermore, 
given the difficulties in constructing group group catalogs 
at low mass halo regime and the associated uncertainties, the
one-to-one correspondence between these groups and dark halos 
is likely to break down, rendering halo mass assignment unreliable.  

We have developed a theoretical framework to compute the PDF of the
host halo mass for a single galaxy, given its neighboring galaxy counts.
We have derived explicit expression for the number of neighboring
galaxies within the cylinder of redshift interval $\pm \Delta z$ and 
transverse comoving distance $r_{\rm p}<r_{\rm p,max}$ in terms of the
HOD model. We include both 1-halo contribution from the same host halo
and 2-halo contribution including chance projection for the neighboring
galaxy counts. We have used the result to obtain the conditional PDF
of the host halo mass given neighboring galaxy counts. 

We compare our analytic model with results from the mock galaxy
catalog, finding reasonable agreements. The PDF of the host halo
mass exhibits complex behavior, with generally two peaks at low- and
high-mass regimes for the case of large neighboring galaxy counts. This is
understood as the effect of chance projection along the
line-of-sight, which can boost neighboring galaxy counts even if the host
halo is a low-mass halo. We find that cylinder radii 
of $\sim 0.5-1h^{-1}{\rm Mpc}$ are optimal for the inference of the
host halo mass, at least for the case of the HOD examined in this
paper.  

This paper serves as a proof-of-concept for our theoretical framework,
and we expect our new approach to have many applications.

%%%%%%%%%%%%%%%%%%%%%%%%%%%%%%%%%%%%%%%%%%%%%%%%%%%%%%%%%%%%
%%%%%%%%%%%%%%%%%%%%%%%%%%%%%%%%%%%%%%%%%%%%%%%%%%%%%%%%%%%%
\section*{Acknowledgments}
%%%%%%%%%%%%%%%%%%%%%%%%%%%%%%%%%%%%%%%%%%%%%%%%%%%%%%%%%%%%%
%%%%%%%%%%%%%%%%%%%%%%%%%%%%%%%%%%%%%%%%%%%%%%%%%%%%%%%%%%%%%
We thank an anonymous referee for useful comments and suggestions.
This work was supported in part by World Premier International
Research Center Initiative (WPI Initiative), MEXT, Japan, and
Grant-in-Aid for Scientific Research from the JSPS (26800093). 
MO acknowledges the hospitality of ASIAA where this work was
partly done.
YTL acknowledges support from the Ministry of Science and Technology 
grant NSC 102-2112-M-001-001-MY3 and the hospitality of Kavli IPMU
where this work was initiated.

%%%%%%%%%%%%%%%%%%%%%%%%%%%%%%%%%%%%%%%%%%%%%%%%%%%%%%%%%%%%%%%%%%%%%%%

\newpage

\begin{appendix}

\newpage

\section{Validity of the log-normal assumption for the PDF of 2-halo
  neighboring galaxy counts}
%%%%%%%%%%%%%%%%%%%%%%%%%%%%%%%%%%%%%%%%%%%%%%%%%%%%%%%%%%%%%%%%%%%%
\begin{figure}
\epsscale{0.95}
\plotone{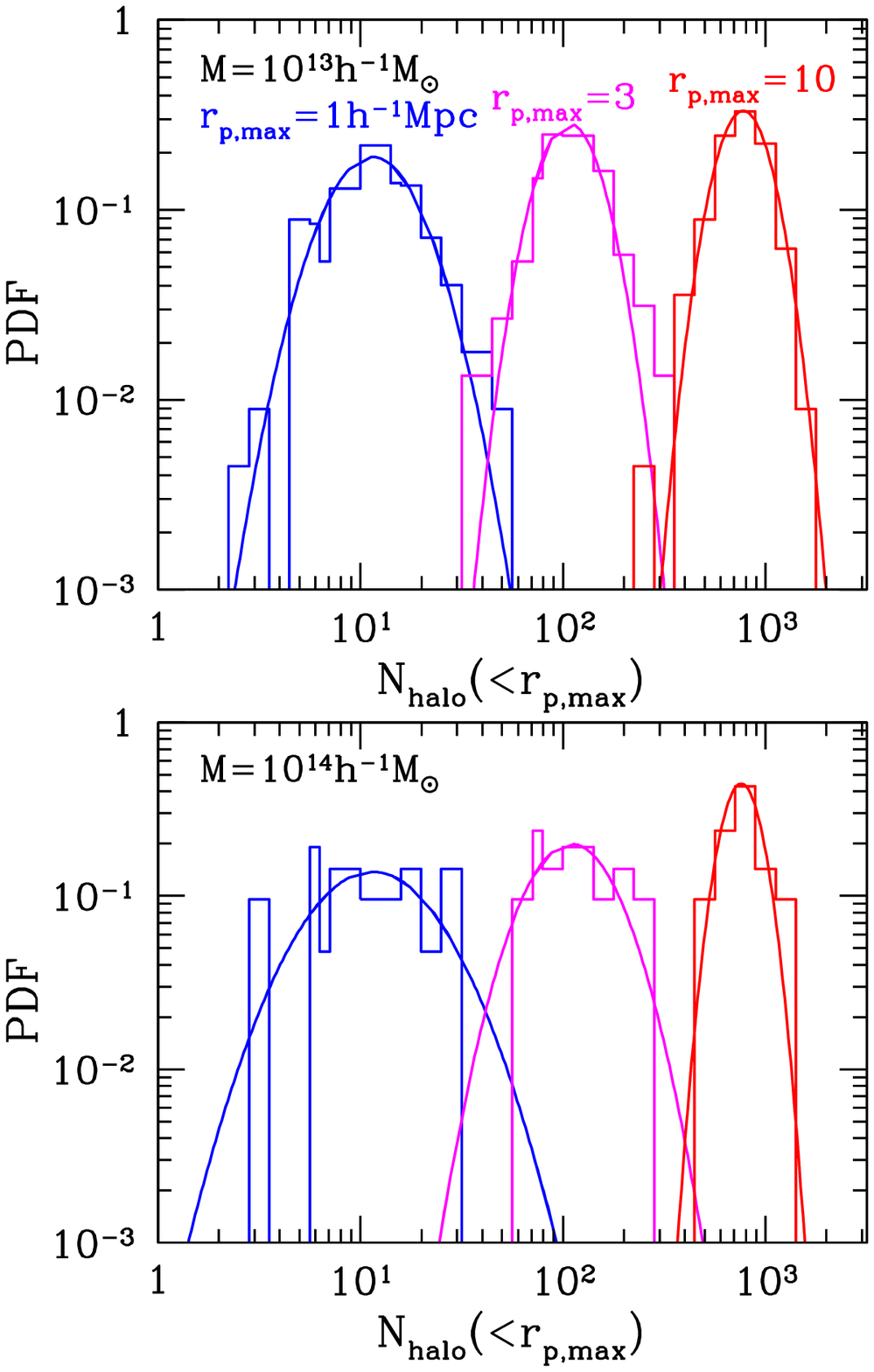}
%\vspace{-4mm}
\caption{ 
The PDFs of number counts of dark matter halos ($M\ge 10^{11}h^{-1}M_\odot$)
around halos with masses $M=10^{13}h^{-1}M_\odot$ ({\it upper panel})
and $10^{14}h^{-1}M_\odot$ ({\it lower panel}). In each panel, we plot
the PDFs for three different maximum transverse separations, $r_{\rm
  p,max}=1$, $3$, and $10h^{-1}{\rm Mpc}$. Histograms show PDFs
measured directly in the Millennium Run simulation (see
Section~\ref{sec:mock} for details), whereas solid lines show best-fit
log-normal distributions.
}
\label{fig:pdf}
\end{figure}
%%%%%%%%%%%%%%%%%%%%%%%%%%%%%%%%%%%%%%%%%%%%%%%%%%%%%%%%%%%%%%%%%%%%

We assumed that the PDF of 2-halo neighboring galaxy counts follows
the log-normal distribution (Equation~\ref{eq:logn}) based on the fact
that the PDF of underlying cosmological density fields is well
described by the log-normal distribution \citep{kayo01}. It is not
clear whether the log-normal distribution can also describe the PDF of
halo number densities which is more closely related to neighboring
galaxy counts. Here we adopt the mock galaxy catalog used in
Section~\ref{sec:mock} to check this log-normal assumption more
directly. 

We consider halo catalogs with masses of $M=10^{13}h^{-1}M_\odot$ and
$10^{14}h^{-1}M_\odot$. For each halo sample, we count the number of
neighboring halos within the maximum transverse distance $r_{\rm
  p,max}$ and the redshift interval $\pm \Delta z=0.02$, as in the
case for our neighboring galaxy counts. We consider all halos with the
masses higher than $10^{11}h^{-1}M_\odot$ which correspond roughly to
the resolution limit of the Millennium Run simulation.

Figure~\ref{fig:pdf} shows the PDFs of the number of neighboring halos
around $M=10^{13}h^{-1}M_\odot$ and $10^{14}h^{-1}M_\odot$ halos, for
the maximum transverse separations of $r_{\rm  p,max}=1$, $3$, and
$10h^{-1}{\rm Mpc}$. We find that the log-normal distribution
generally fits very well. A possible exception is the case for small
$r_{\rm   p,max}$ and the halo mass of $M=10^{14}h^{-1}M_\odot$ for
which effects of halo exclusion and non-Gaussian error are most
prominent, but even in this case the log-normal distribution provides
a reasonably good approximation. Given the log-normal PDF of halo
number counts, we expect neighboring galaxy counts also obey the
long-normal distribution.
\end{appendix}

\end{document}